# An Appropriate Sensor Distribution Technique in Wireless Sensor Networks

Malathi Balaji,
*Research Scholar,*
*Madurai Kamaraj University,*
*Madurai, India*
bmalathisai@gmail.com

Dr. Alagarsamy,
*Professor,*
*Madurai Kamaraj University,*
*Madurai, India*
alagarsamy_mku@gmail.com

Dr. Veeramani. V,
*Assistant Professor,*
*College of Applied Sciences*
*–Salalah, Oman*
veeramani.sal@cas.edu.om

Dr.RD. Balaji,
*Assistant Professor,*
*College of Applied Sciences –*
*Salalah, Oman*
balaji.sal@cas.edu.om

*Abstract* - Wireless Sensor Network (WSN) is pertinent to many applications with varied network parameters. Sensor node placement in the application region whether it is indoor or outdoor is a major task as well as plays very remarkable role in the network performance. Node placement is carried out according to the region where it is applied, either deterministic or non deterministic. Because of the need for different sensing probability or detection probability, same approach of sensor placement may not be suited for all the applications. Some of the applications are well formed and give better performance with the uniform distribution of sensors but few need intense distribution of nodes in particular sensitive places especially those applications meant for intrusion detection. An application which needs high level intrusion detection and a suitable sensor distribution methodology, known as Half-Normal Distribution (Half-Gaussian) based deployment, have been set forth in this paper. We have also discussed the theoretical comparison for detection probability with the uniform distribution in terms of number of sensor nodes.

*Keywords - Wireless Sensor Network, Distribution and deployment, Half-Normal Distribution, Intrusion Detection, Detection probability*

## I. INTRODUCTION

The world has been fascinated by automation in many day to day aspects. The Information and Communication Technology (ICT) has brought many innovations in the form of hardware and software. One of the major contributions which is dominantly used in many intelligent applications from small scale to large scale is WSN. It involves many real-time applications including security, health care, intelligent monitoring and environmental studies. Worldwide in many aspects of WSN development, researches are continuing; including localization, node deployment, routing, scalability, mobility, intruder tracking, sensitivity, coverage and lifetime, among them the sensor node deployment plays a pivotal role. Deployment of sensors in the Region of Interest (RoI) is the first task to be done for any WSN and it is purely application specific, in-door or out-door, smaller and accessible region or larger and inaccessible region. Few important parameters are to be noted before and after deploying the WSN in the RoI, including network and sensing coverage, density of nodes and the deployment model [1]. The communication among sensor nodes and to the sink should be streamlined, so that no area is left uncovered and no sensor is left disconnected. Sensing range of the nodes varies according to the sensor type and should be carefully chosen for any specific requirement. The number of nodes per square meter is the density of nodes; uniform density in the RoI is expected to be maintained for better coverage. There are different deployment methodologies under the major categories of deterministic, non-deterministic and semi-deterministic [2]. Grid placement is deterministic where the nodes are placed strictly on the line of a grid, which is practically less feasible. Biased random placement is semi-deterministic where the area chosen is deterministic but the distribution in that area is random which is non-deterministic. Non-deterministic (stochastic) placement can either be simple diffusion (e.g. dropped from the airplane) or random (uniform) placement, which are very realistic and easy to use [3]. Few deployments are probability distribution based and few are distribution-free approaches. Though the words "distribution and deployment" are interchangeably used by many researchers to imply the same meaning for sensor placement and it is not wrong too, they have notably different meanings. In case of WSN, distribution means spreading out the sensors on the area of interest based on some rule, where as the deployment means more than that, i.e. installation of sensors in a particular place with proper configuration and with all needed customization.

The scope of our paper is limited to sensor node deployment which may enhance the intrusion detection of human, animals or vehicles; here we introduce a new distribution based sensor deployment, known to be Half- Normal Distribution based deployment, which is meant for a special application scenario. Half-Normal distribution is a special case of Folded- Normal distribution with mean value zero. The remaining paper is organized as follows. The next section lists some of the similar work and followed by the explanation of Half-Normal distribution method and theoretical proof for its performance with future work and conclusion.

## II. RELATED WORK

In WSN, different methodologies are followed for sensor deployment in out-door and hard –terrain regions. Compared to random deployment, the probabilistic distribution based sensor deployments in hard-terrain application regions are yielding better performance in case of coverage, lifetime of WSN, stability of protocols [4] and intruder detection.

Y.Wang et.al [5] have proved that the Gaussian distribution based sensor deployment in WSN improves the detection probability in protecting a target region. This method provides differentiated detection probability and outperforms the uniform distribution. They have showed theoretically and with simulation that the performance of Gaussian distribution in single sensing and multiple sensing models, with the necessary network parameters including number of deployed sensors.

Omar and Alaa [6] have compared uniform, normal and mixed distributions with suitable simulations and concluded that the normal distribution is showing better performance in terms of intruder detection, average number of transmitted control bits and number of hops and mixed distribution is better for reduced end-to-end delay. It has been proved by [7] that in the multi-terrain environment, the power efficient routing heuristics for maximizing network life time (CMAX, OML) are enhanced by the Normal, Poisson, Uniform and Chi-square distributions in different combinations.

Collaborative beamforming [8] has been proposed which gives better performance in the Gaussian distributed WSN in terms of wider mainlobe and lower chance of large sidelobes when compared to uniform distribution. Q.Yu, et.al [9] have proved that even when the intruder is destroying the sensors in his intruding path towards the target, if sensors are deployed in Gaussian distribution, the detection probability is higher than the uniform distribution. From the paper [10], it is known that the exponential distribution of sensor nodes towards the sink and hybrid routing improves the lifetime of the network as well as reduces the energy hole problem and its performance is better while comparing with uniform distribution.

Shaila.K et.al [11] have proposed a probabilistic model for single and multiple sensing Intrusion Detection in 2-D and 3-D homogeneous WSN. They have used the clustering technique where only the cluster head communicates with the sink, thus reducing the number of sensor nodes participating in intrusion detection, this in turn enhances the lifetime of the network. The nodes are redistributed randomly after a certain period of time in order to reduce the early death of nodes. They have proved with simulation that their method of deployment has achieved minimum energy consumption both in 2-D and 3-D network.

In the next section, we discuss the details of the proposed approach.

## III. HALF –NORMAL DISTRIBUTION BASED SENSOR NODE DEPLOYMENT:

Some special applications have got lengthy locations from few meters to more kilometers which may need monitoring both the sides of it or in one side and not continuous too, which is sometimes known as barrier coverage. A sample application of Half-Normal distribution is road monitoring. Though many established techniques are being used to guarantee the road safety, the researches are still continuing. In many countries, the desert roads and village roads are not crowded but animal crossing is more common. Since it is not crowded the vehicles are driven faster and this causes danger to the lives of both humans and animals. These road sides can be protected with WSN in the half normal distribution of sensors as shown in figure 1. More number of sensors near the roads than the region away from the roads with a suitable localization technique is best suited to warn the drivers to go slow and to protect the animals. When the animal movement is towards the road, the sensors pass the information to the control unit and through that the Road Side Units (RSU) [12] give the warning light to the drivers or the message is sent directly to the cars coming towards the place through the On Board Unit (OBU). In case of military applications, to enhance the intruder detection near the highly sensitive regions, Half-Normal distribution based deployment of sensor nodes would be an optimal choice. The proposed distribution method will be appropriate to improve the performance of WSN not only in terms of intrusion detection, but to enhance coverage and lifetime of the network and it is suitable for hard – terrain regions like hilly and mountainous regions also.

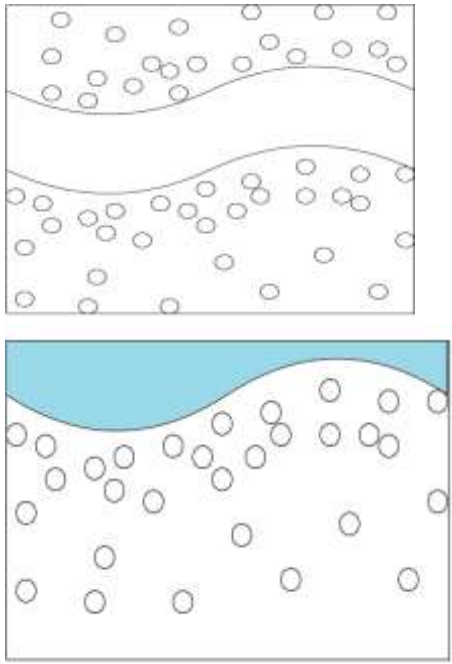

Fig. 1 Half-Normal Distribution

The (truncated) Gaussian distribution is applicable for any closed targets or region to provide better intrusion detection and the same may not be well suited for such above mentioned applications. Uniform distribution is useful for spreading the sensors uniformly, which is good when we need same sensing probability but if the application needs differentiated detection probability, uniform distribution may not satisfy the requirement. The Half- Gaussian distribution [13 and 14], which is a special case of the folded Gaussian distribution [15], can be adopted for such scenarios. The folded Gaussian distribution is a probability distribution related to the normal distribution, where the probability mass to the left of X=0 is folded over by taking the absolute value.

Let X follows the ordinary Gaussian distribution N(0, $\sigma^2$), then Y=|X| follows a Half- Gaussian distribution, which is a fold at the mean of an ordinary normal distribution with mean zero. With the $\sigma$ parameterization of the normal distribution, the probability density function (PDF) of the Half-Normal is given by,

$$f_Y(y;\sigma) = \frac{\sqrt{2}}{\sigma\sqrt{\pi}} \exp\left(-\frac{y^2}{2\sigma^2}\right) \quad y > 0 \quad \text{Where}$$

$$E[Y] = \mu = \frac{\sigma\sqrt{2}}{\sqrt{\pi}} \quad (1)$$

The PDF and CDF curves of the half –normal distribution are given in the figure 2.

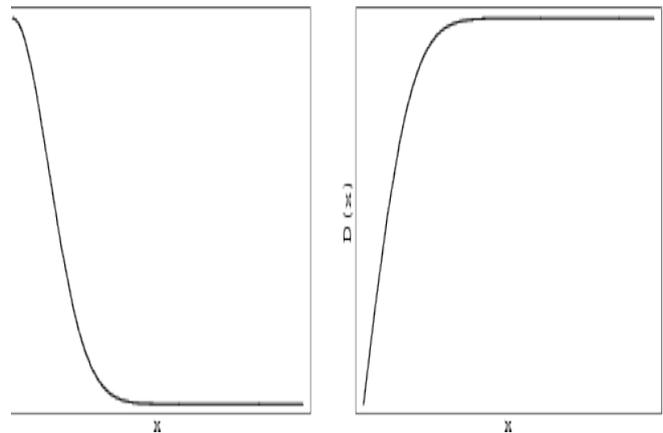

Fig. 2 PDF and CDF of Half-Normal Distribution (Courtesy: archive.lib.msu.edu)

The performance algorithms which were proven for normal distribution are suited for the half- normal distribution with the positive values. The normal distribution truncated at 0 and [16] folded (half) normal distribution give same results for a mean – 0 normal distribution, because any distribution that is symmetric about 0 and has 0 probability of being 0. If the mean is nonzero value, the distribution is not symmetric and both the distribution methods do not give the same result. It is an added proof from [17] for the Half-Normal distribution to have absolute value with the Stein's method of density approach. It is given that, a random variable X with values in [0, ∞] has the Half-Normal distribution μ if and only if E [f´(x)] = E [x f(x)] –f(0)$\frac{\sqrt{2}}{\sqrt{\pi}}$ for all functions f: [0,∞]→ R, which are absolutely continuous on every compact sub-interval of [0, ∞] such that E |f´ (y)| < ∞.

For the sensor deployment, 2D Half-Normal Distribution has to be adopted. The probability density function is given as [18],

$$f_{X,Y}(x, y) = 2[\pi\sigma_1\sigma_2\sqrt{(1-\rho^2)}]^{-1} * [\exp\{-(\frac{x}{\sigma_1})^2 + (\frac{y}{\sigma_2})^2 /2(1-\rho^2)\}]\cosh[\frac{\rho xy}{(1-\rho^2)\sigma_1\sigma_2}]$$

$$0<x, 0<y \qquad (2)$$

The X and Y each have Half- Normal distribution in two dimensions with variance $(1-\rho^2)$, where $\rho$ is the correlation between X and Y. The definition of correlation is $\rho_{X,Y} = \frac{Cov[X,Y]}{\sigma_1\sigma_2}$, when we consider X and Y are independent, $\rho = 0$. We assume the same standard deviation along x and y dimensions for sensor node deployment $\sigma_1 = \sigma_2 = \sigma$, so the PDF is

$$f_{X,Y}(x, y) = 2[\pi\sigma^2]^{-1} * [\exp\{\frac{-(x^2+y^2)}{\sigma^2}/2\}] \quad (2.a)$$

Any WSN is deployed with an intended requirement, without which the proposal is incomplete. Here we aim to alleviate the problem of intrusion detection with the appropriate node deployment. The following section discusses the theoretical proof for the furtherance of intrusion detection with the Half-Normal distributed WSN.

### A. Terminologies:

Few terminologies used in this paper are listed below:

R - Region of Interest (R and RoI are interchangeably used in this paper)
N - Total number of sensors in R
S - Starting position of Intruder
d - Intruder moving distance
D - Max distance permitted to travel before reaching the target region inside the R
r - Sensing range of a sensor
A - Intrusion detection area
$P_d$ - Probability of the intruder to be detected by at least one sensor node in R

### B. System model and Assumptions:

Sensing model: For simplicity we consider Deterministic Boolean sensing model for better network coverage. [19] Let R is the Region of Interest (RoI), r is the sensor's sensing range, and N is the total number of sensors. The probability of the intruder to be detected by any sensor is p=$\frac{\pi r^2}{R}$, so the undetected by any sensor probability is 1 - p; the probability for the intruder not to be sensed by any of the N sensors is $P_n = (1 - p)^N$. Thus the probability of the intruder to be detected by at least one sensor node among N nodes is

$$P_d = 1 - (1 – p)^N \qquad (3)$$

Detection model: Here we consider only single sensing model, the intruder is detected by any single sensor.

We assume the straight line motion of the intruder and single intruder detection, where the intruder can be a person, animal or vehicle. The movement detecting nodes are employed in the RoI [20] which are cost effective. The assumption includes that the nodes use multi-hop communication to the smart sink which does the data processing and all the nodes identify their neighbors with full network coverage.

### IV. INTRUSION DETECTION IN SINGLE SENSING MODEL

If the intruder is detected before D (d ≤ D), then the system performance is good as shown in the following figure 3. From S, the movement of the intruder is calculated with Cartesian system, [9 and 5] which includes a rectangular area and two half disk areas as per the intruder movement and sensing range. The area of intrusion detection is given by

$$A = 2Dr + (\frac{\pi r^2}{2}) + (\frac{\pi r^2}{2}) \qquad (4)$$

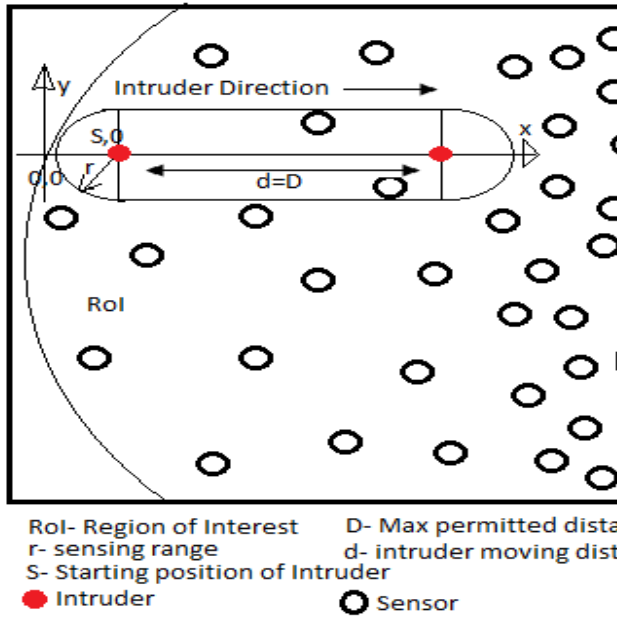

Rol- Region of Interest    D- Max permitted dista
r- sensing range            d- intruder moving dist
S- Starting position of Intruder
● Intruder                  ○ Sensor

Fig. 3 Intrusion detection in a Half-Normal Distributed WSN

The probability of the presence of a sensor node in the above mentioned rectangular area is

$$P_{rect} = \int_{S-d}^{S} \int_{-r}^{r} f_{xy}(\sigma) dy\, dx \quad (5)$$

where the function is the Half-Normal Distribution PDF.

The probability of a sensor node to be deployed in the first half disk area is

$$P_{left} = \int_{S-d-r}^{S-d} \int_{-\sqrt{r^2-(x-S+D)^2}}^{\sqrt{r^2-(x-S+D)^2}} f_{xy}(\sigma) dy\, dx \quad (6)$$

The probability of a sensor node to be in the second half disc area is

$$P_{right} = \int_{S}^{S+r} \int_{-\sqrt{r^2-(x-S)^2}}^{\sqrt{r^2-(x-S)^2}} f_{xy}(\sigma) dy\, dx \quad (7)$$

Thus the probability of sensor to be deployed in the intrusion detection area A is

$$P_{total} = P_{rect} + P_{left} + P_{right} \quad (8)$$

As per the Boolean sensing model in equation (3), the probability of the intruder to be detected at least by one sensor node among N nodes is,

$$P_d = 1 - (1 - P_{total})^N \quad (9)$$

Equation 9 can be written as,

$$P_d = 1 - (1 - (\int_{S-d}^{S} \int_{-r}^{r} f_{xy}(\sigma) dy\, dx + \int_{S-d-r}^{S-d} \int_{-\sqrt{r^2-(x-S+D)^2}}^{\sqrt{r^2-(x-S+D)^2}} f_{xy}(\sigma) dy\, dx + \int_{S}^{S+r} \int_{-\sqrt{r^2-(x-S)^2}}^{\sqrt{r^2-(x-S)^2}} f_{xy}(\sigma) dy\, dx))^N \quad (10)$$

It has been theoretically shown that the intruder will be detected during his (it's) movement towards the target region at least with single sensor before reaching the maximum permitted distance D.

The following figure 4 shows the theoretical comparison graph between Half-Normal Distribution and Uniform Distribution in terms of number of sensors deployed. When sensor count increases, the detection probability increases to the threshold. An important thing to be noted here is when the intruder starts from the boundary of the RoI (S=0), the Uniform distribution is better than the Half-Normal because comparatively number of sensor nodes would be more in the former method as the sensors are evenly distributed. But when the intruder enters the RoI suddenly close to the target region, say S=5, (For eg: dropped from helicopter in the military field, animals enter near the road from any unexpected sides), the uniform distribution will remain same in detection probability, but the Half-Normal method would perform faster as the number of sensors are more and sensing probability is very high. Thus the proposed Half-Normal distribution will provide better detection probability in critical situations.

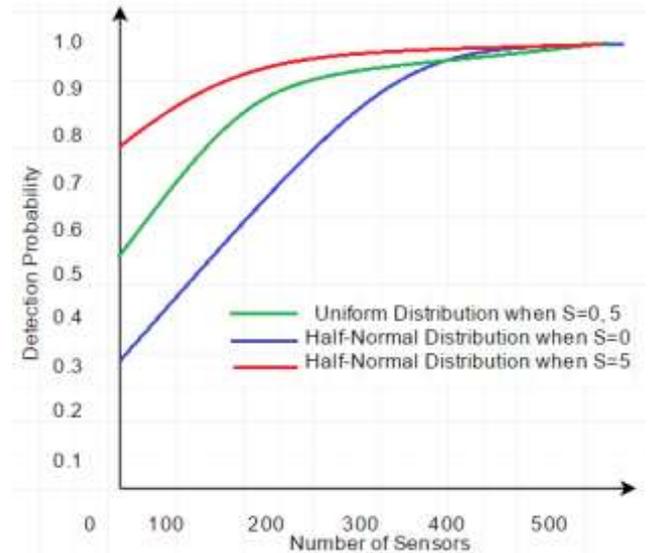

Fig. 4 Comparison graph of Half-Normal distribution and normal distribution in terms of Detection Probability Vs Number of sensors.

## V. CONCLUSION AND FUTURE WORK

Wireless Sensor Network is a growing technology which involves many research areas in its development and application. In this paper we have chosen the deployment of sensor nodes for our research, which is application specific and plays a pivotal role in the performance of the WSN. The probabilistic Half-Normal distribution based deployment approach is discussed in this paper with its relevant application domain and theoretical proof for achieving better intrusion detection.

The simulation of our proposed Half-Normal distribution based sensor deployment method is yet to be done. To detect a single intruder, it may be ineffectual to involve all the sensors in the location as the lives of the sensors are affected as well as the network traffic increases. Instead when the nodes are formed into clusters with a Cluster Head (CH) for each cluster, only the CH will communicate the sensed data to the sink, hence reducing the traffic and improving the life time of the sensors [11]. So the concept of clustering can be introduced in our proposal for pretty good performance of the network. The sensors are deployed in RoI from airplane if the region is tricky for manual deployment. This method is not always commendable due to the unexpected network disturbances with natural phenomena, like nodes gathered in a location and no coverage in other areas because of heavy wind. This kind of issues can be handled with optimization and redeployment after deployment with third party involvement (actuators, robots etc) [21]. Usage of mobile nodes may cut the issue to some extent, but they consume more battery for movement. We would like to alleviate the above mentioned problem in our future work.

## REFERENCES

[1] Majdi Mansouri, Ahmad Sardouk, Leila Merghem-Boulahia," Factors that may influence the performance of wireless sensor networks", *www.intechopen.com, ISBN 978-953-307-261-6, 2010*

[2] Charalambos Sergiou and Vasos Vassiliou, "Efficient Node Placement for Congestion Control in Wireless Sensor Networks", *European Union Project GINSENG funded under the FP7 Program (FP7/2007-2013)*

[3] Malathi, et.al, "Analysis of Different Sensor Deployment Strategies in Wireless Sensor Network" *International Journal of Advanced Research in Computer Science and Software Engineering, Volume 5, Issue 10, October-2015 ISSN: 2277 128X*

[4] Saleh H. Al-Sharaeh, et.al, *"*Multi-Dimensional Poisson Distribution Heuristic for Maximum Lifetime Routing in Wireless Sensor Network", *World Applied Sciences Journal 5 (2): 119-131, 2008 ISSN 1818-4952*

[5] Yun Wang, Weihuang Fu, and Dharma P. Agrawal, " Gaussian versus Uniform Distribution for Intrusion Detection in Wireless Sensor Networks", *IEEE Transactions on Parallel And Distributed Systems, Vol. 24, No. 2, February 2013.*

[6] Omar Said, Alaa Elnashar, "Optimizing Sensors Distribution for Enhancing WSN Intrusion Detection Probability in Euclidian's Space", *International Journal of Computer Applications (0975 – 8887) Volume 105 – No. 15, November 2014*

[7] Fatima M. Osman and Saleh H. Al-Sharaeh*," Hetrogeneous Multi-Deployment Strategy Effect on Maximizing the Lifetime Routing in Wireless Sensor Network", Middle-East Journal of Scientific Research 13 (6): 749-759, 2013 ISSN 1990-9233*

[8] Mohammed F. A. Ahmed, Sergiy A. Vorobyov, " Collaborative Beamforming for Wireless Sensor Networks with Gaussian Distributed Sensor Nodes", *IEEE Transactions on Wireless Communications, Vol. 8, No. 2, February 2009*

[9] Qixiang Yu, Zhenxing Luo, and Paul Min," Intrusion Detection in Wireless Sensor Networks for Destructive Intruders", *Proceedings of APSIPA Annual Summit and Conference 2015*

[10] Aruna Pathak, Zaheeruddin and D.K. Lobiyal, "Maximization the Lifetime of Wireless sensor network by minimizing Energy hole problem with Exponential node distribution and hybrid routing", *Engineering and Systems (SCES), March 2012 Students Conference.*

[11] Shaila K, et.al, "Probabilistic Model for Single and Multi-Sensing Intrusion Detection in Wireless Sensor Networks", *IOSR Journal of Computer Engineering (IOSR-JCE) e-ISSN: 2278-0661, p-ISSN: 2278-8727Volume 16, Issue 1, Ver. IX (Feb. 2014), PP 51-66 www.iosrjournals.org*

[12] Mohammad Jalil Piran, "Vehicular Ad Hoc and Sensor Networks; Principles and Challenges*", International Journal of Ad hoc, Sensor & Ubiquitous Computing (IJASUC) Vol.2, No.2, June 2011*

[13]https://en.wikipedia.org/wiki/Half-Normal_distribution

[14] M.Ahsanullah et.al., "Normal and Student's t Distributions and their Applications", *Atlantis Studies in Probability and Statistics 4, Atlantis Press and the authors 2014*

[15]https://en.wikipedia.org/wiki/Folded_normal_distribution

[16]http://stats.stackexchange.com/questions/16089/is-sampling-from-a-folded-normal-distribution-equivalent-to-sampling-from-a-norm

[17] Christian Dobler, " Stein's method for the Half-Normal Distribution with applications to limit theorems related to simple random walk", *Arxiv:1303.4592v2 [Math.PR] 2013*

[18] Samuel Kotz, N. Balakrishnan, Norman L. Johnson, "Continuous Multivariate Distributions, Models and Applications", *https://books.google.com/books?isbn=0471654035, 2004*

[19] Ashraf Hossain, P. K. Biswas, S. Chakrabarti, "Sensing Models and Its Impact on Network Coverage in Wireless Sensor Network", *2008 IEEE Region 10 Colloquium and the Third ICIIS, Kharagpur, INDIA December 8-10.*

[20] http://www.alibaba.com/showroom/motion-sensors-prices.html

[21] Hanen Idoudi, et.al, "Robots-Assisted Redeployment in Wireless Sensor Networks" *Journal of Networking Technology, Dline, Vol. 3, No. 1, March 2012*